\documentclass[useAMS,usenatbib]{mnras}
\usepackage{graphicx,psfig,cite,epsf}  
\usepackage{times}
\usepackage{subfigure}
\usepackage{textcomp}
\usepackage{multirow}
\usepackage{amsmath}	% Advanced maths commands
\usepackage{amssymb}	% Extra maths symbols
\usepackage{gensymb}
\usepackage{array}

\def\apjl{ApJ}                % Astrophysical Journal, Letters
\def\apjs{ApJS}               % Astrophysical Journal, Supplement
\def\ao{Appl.~Opt.}           % Applied Optics
\def\aap{A\&A}                % Astronomy and Astrophysics
\def\xmm{{\it XMM-Newton}}
\def\nus{{\it NuSTAR}}
\def\chandra{{\it Chandra}}

\title[The two ULXs in NGC 925]{The two Ultraluminous X-ray sources in the galaxy NGC 925}

\author[Pintore F. et al.]{F. Pintore$^1$\thanks{E-mail: pintore@iasf-milano.inaf.it}, L. Zampieri$^2$, S. Mereghetti$^1$, A. Wolter$^3$, G. Rodr\'iguez$^4$, G. L. Israel$^4$, \newauthor P. Esposito$^5$, S. Paiano$^2$, G. Trinchieri$^3$, P. Ochner$^2$\\
 $^1$ INAF - IASF Milano, Via E. Bassini 15, I-20133 Milano, Italy\\
 $^2$ INAF - Osservatorio Astronomico di Padova, Vicolo dell'Osservatorio 5, I-35122 Padova, Italy\\
 $^3$ INAF - Osservatorio Astronomico di Brera, Via Brera 28, I-20121 Milano, Italy; \\
 $^4$ INAF - Osservatorio Astronomico di Roma, Via Frascati 44, I-00040, Monteporzio Catone, Italy; \\ 
 $^5$ Anton Pannekoek Institute for Astronomy, University of Amsterdam, Science Park 904, 1098 XH Amsterdam, The Netherlands }

\date{Accepted  . Received  ;}

\pubyear{2017}

\begin{document}

\maketitle

\begin{abstract}

NGC 925 ULX-1 and ULX-2 are two ultraluminous X-ray sources in the galaxy NGC 925, at a distance of 8.5 Mpc. For the first time, we analyzed high quality, simultaneous \xmm\ and \nus\ data of both sources. 
Although at a first glance ULX-1 resembles an intermediate mass black hole candidate (IMBH) because of its high X-ray luminosity  ($(2$$-$$4)\times10^{40}$ erg s$^{-1}$) and its spectral/temporal features, a closer inspection shows that its properties are more similar to those of a typical super-Eddington accreting stellar black hole and we classify it as a `broadened disc' ultraluminous X-ray source. Based on the  physical interpretation of this spectral state, we suggest that ULX-1 is seen at small inclination angles, possibly through the evacuated cone of a powerful wind originating in the accretion disc. The spectral classification of ULX-2 is less certain, but we disfavour an IMBH accreting at sub-Eddington rates as none of its spectral/temporal properties can be associated to either the soft or hard state of Galactic accreting black hole binaries.

\end{abstract}

\section{Introduction}
Ultraluminous X-ray sources (ULXs) are extragalactic, point-like objects characterized by very high X-ray luminosities in the range $10^{39}-10^{42}$ erg s$^{-1}$ (e.g. \citealt{fabbiano89,fengsoria11,kaaret17}).
Observational evidences suggest that ULXs are accreting X-ray binaries (XRBs) with massive donors (e.g. \citealt{liu13,motch14}). 
The ULX luminosities can be produced from super-Eddington accretion on stellar-origin black holes (BHs) that could be similar to the Galactic ones (e.g. \citealt{gladstone09,sutton13,middleton15}) or moderately more massive (e.g. \citealt{zampieri09}). Alternatively, a possibility can be sub-Eddington accretion onto intermediate mass BHs (IMBHs; e.g. \citealt{colbert99,madau01,portegies04,miller02}). Another possibility is super-Eddington accretion onto neutron stars (NSs; e.g. \citealt{bachetti13,israel16a}).
The combination of spectral and temporal properties of ULXs should in principle allow the distinction between the nature and mass of the different accretors. 
In fact, IMBHs accreting sub-Eddington should present the hallmarks of the accreting states of Galactic BH binary systems (i.e. {\it hard} and {\it soft} states; \citealt{mcclintock06}), while super-Eddington accreting stellar BHs and NSs would more likely show the features of the {\it ultraluminous state} \citep[see e.g.][]{roberts07,gladstone09}. The latter is characterized spectrally by a curvature at energies of $2-5$ keV, often associated to a thermal soft excess below $\sim$$ 0.5$ keV (e.g. \citealt{gladstone09, bachetti13,caballero13, rana14, walton13,walton14}), and temporally by random short-term variability \citep[e.g.][]{heil09, sutton13,pintore14}. { The high energy curvature may arise either from a cold, optically-thick corona lying above an accretion disc (e.g. \citealt{poutanen07, gladstone09, pintore12}), or i) from the innermost region of an advection-dominated accretion disc \citep[e.g.][]{mizuno07}, ii) from the reprocessing of hard photons by an optically-thick outflow, iii) from a combination of these effects.} The soft component may instead originate from the photosphere of a radiatively-driven and clumpy wind ejected from the accretion disc \citep[e.g.][]{pinto16} when the local luminosity at the surface overcomes the Eddington limit, as expected for super-Eddington accretion rates \citep[e.g.][]{poutanen07,ohsuga09,takeuchi13}. These powerful outflows may also be responsible for the unpredictable ULX short-term variability \citep[e.g.][]{middleton15}. 

\begin{figure*}
\center
		\includegraphics[width=15cm]{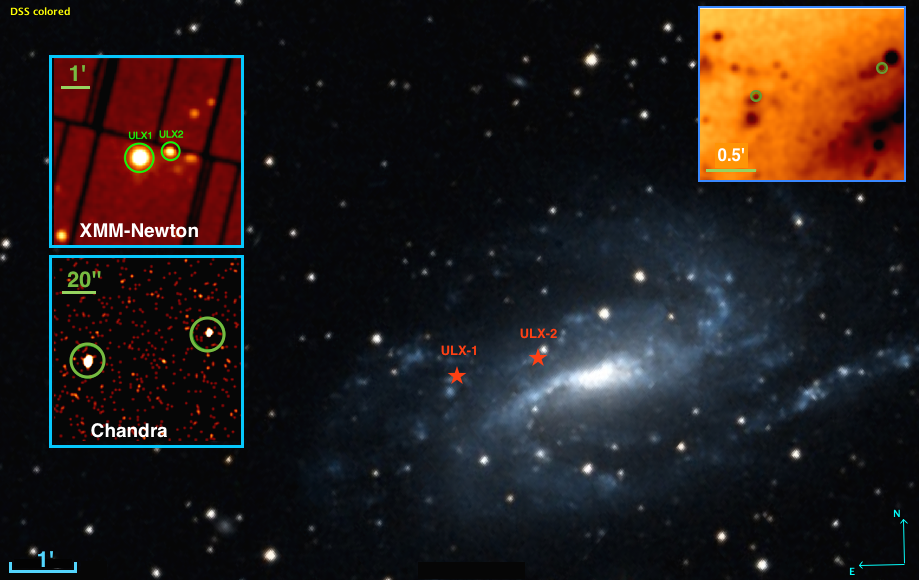}
   \caption{DSS image of the galaxy NGC 925, where the red stars indicate the positions of ULX-1 and ULX-2. Insets: \xmm/EPIC-pn (left-top), \chandra/ACIS-S (left-bottom) and DSS images (right). The images have different scales.}  
        \label{allimage}
\end{figure*}

The discovery of four pulsating ULXs (PULXs) in M82 ULX-2, NGC 5907 X-1, NGC 7793 P13 and NGC 300 ULX-1 (\citealt{bachetti14, israel16a}a; \citealt{israel16b}b; \citealt{fuerst16a}; \citealt{fuerst17}; \citealt{carpano18a}) demonstrated that ULXs can contain NSs, indicating clearly that the ULX population does not host only BHs. Furthermore, the PULXs enlarged our view of the accretion physics in ULXs and, in general, of the extreme mechanisms that can power accreting NSs.
{ The four PULXs currently represent the most super-Eddington accreting sources. Their spectral and temporal properties are quite similar to other well studied ULXs, even though they may have a harder emission \citep[e.g.][]{pintore17,walton18}, that could originate in a post-shock region of the accretion column above the NS surface. Magnetic fields of $\sim 10^{14}$ G or larger can produce high isotropic luminosity  in the accretion column (up to 2-3 orders of magnitudine higher than the ordinary Thomson Eddington limit), as shown by semi-analytical calculations and 2D radiation-hydro-simulations of accretion onto NSs with high magnetic fields \citep{mushtukov15,mushtukov17, kawashima16}.}

The ULX population has continuously increased during the years reaching several hundreds known examples, although only a few of them have high-enough quality X-ray data to perform deep investigations. 
Here we report on two ULXs in the galaxy NGC 925 (SAB(s)d, $D\sim$ 8.5 Mpc; Figure~\ref{allimage}). These are NGC 925 ULX-1 (CXO J022727+333443, ULX-1 hereafter) and ULX-2 (CXO J022721+333500, ULX-2 hereafter), both listed in the \textit{Chandra} ULX catalogue \citep{swartz11}. ULX-1 is located in a spiral arm, while ULX-2 is at $\sim80''$ from ULX-1. Both sources were observed 18 times by \textit{Swift/XRT}, which showed that ULX-1 presented flux variations up to a factor of 3, with a peak 0.3-10 keV luminosity of $\sim4\times10^{40}$ erg s$^{-1}$ (ULX-2 was instead below the XRT threshold). \citet{heida16} analyzed their infrared emission and estimated that the ULX-1 donor star cannot be a supergiant star of spectral type F or later, while for ULX-2 a red supergiant star may be a possible companion. 

{ From a preliminary analysis of the only public, short, \chandra\ observation of the two sources, we found that they showed hard spectra and high short-term temporal variability. Since the combination of these properties might be an indication of an accreting IMBH, in this work we investigate further the nature of both sources using new, simultaneous, high-quality \xmm\ and \nus\ observations.}

\section{data reduction}
\label{data_reduction}

\subsection{XMM-Newton}

An \xmm\ observation of ULX-1 and ULX-2 was taken on 2017-01-18, for a total exposure time of $\sim50$ ks. We extracted the data obtained with the EPIC-pn and the two EPIC-MOS cameras, both of them operated in full-frame mode. 
We reduced the data with the SAS v15.0.0 software, selecting single- and double-pixel events ({\sc pattern}$\leq$4), and single- and multiple-pixel events ({\sc pattern}$\leq$12), for pn and the MOS, respectively. We cleaned the data removing high particle background time intervals and resulting in net exposure times of $\sim32$ ks and $\sim42$ ks in the pn and MOS, respectively.

For the spectral and timing analysis, we extracted the background data from circular regions of 60$''$ radius, free of sources and close to the two ULXs. The ULX data were extracted from circular regions of 35$''$ and 22$''$ radii for ULX-1 and ULX-2, respectively (see Figure~\ref{allimage}). The smaller ULX-2 radius was chosen because the source was very close to a CCD gap.
We derived a total of $\sim$16920/13512 and $\sim$1643/2420 net counts in EPIC-pn/MOS for ULX-1 and ULX-2, respectively.

\subsection{NuSTAR}
We also obtained a $\sim$42 ks \nus\ observation which started $\sim$20 minutes before the \xmm\ one, thus broadly overlapping with it.
The \nus\ data were reduced using the standard pipeline, based on {\sc NUSTARDAS} (the \nus\ {\it  Data Analysis Software} v1.3.0) in the HEASOFT {\sc ftools} v6.16 { and CALDB version 20180312}. We obtained cleaned event files by applying standard procedures. We extracted the ULX-1 data, selecting a circular region of 50$''$ radius centered on the source. The background was chosen from a nearby circular region, free of sources, of 80$''$ radius.
We obtained a total of $\sim2300$ net counts in the sum of the data from the FMPA and FMPB instruments. { Although ULX-2 was very faint for \nus\ and also close to ULX-1, a circular region of $30''$ was used to extract ULX-2 data in order to avoid strong ULX-1 contamination.}

\subsection{Chandra}
We analyzed an archival \chandra/ACIS-S observation (Obs.ID. = 7104) of 2005-11-23 with an exposure time of $\sim$$2.2$ ks. \chandra\ data were reduced with {\sc ciao} v.4.9 and calibration files CALDB v.4.7.6.  
The source events were chosen from circular regions of 3$''$ radius (adequate for off-axis position of the sources; see Figure~\ref{allimage}), while the background events were selected in a close circular region of 15$''$ radius.
We extracted the source spectra with the {\sc ciao} task SPECEXTRACT, which generates the appropriate response and auxiliary files for the spectral analysis. 
\newline
\newline
\newline
The NuSTAR spectra were grouped with at least 100 counts per bin while the \xmm\ and \chandra\ spectra were grouped with 25 counts per bin, so that minimum $\chi^2$ fitting techniques could be used. 

Model fitting was carried out using XSPEC v.12.8.2 \citep{arnaud96}. 
The spectra of the pn and the two MOS cameras, and (when available) of the \nus\ detectors were fit together. { A multiplicative factor was included to account for possible inter-calibration uncertainties that, as expected, we found smaller than 12\% \citep[e.g.][]{madsen17}}. \chandra\ was instead analyzed singularly as it was not simultaneous with the other two datasets.
For the spectral fits, we considered the 0.3--10 keV energy range for \chandra\ and EPIC data, and 3--70 keV energy range for \nus\ data.

\subsection{Optical data}
On 2018 March 25, starting at 18:44:31.3 UTC, we took two images (300s+200s) of the field of ULX-1/ULX-2 in the H$_{\alpha}$ band with with the 1.8-m Copernico Telescope at Cima Ekar (Asiago, Italy). They were reduced and analyzed using standard software and procedures (bias and flat field subtraction, astrometric calibration). 

We report in this section that both ULX-1 and ULX-2 are surrounded by a diffuse emission in H$_{\alpha}$, although the quality of the two images does not allow us to investigate the optical emission in deeper detail.

\begin{figure}
\center
		\includegraphics[angle=270,width=8.3cm]{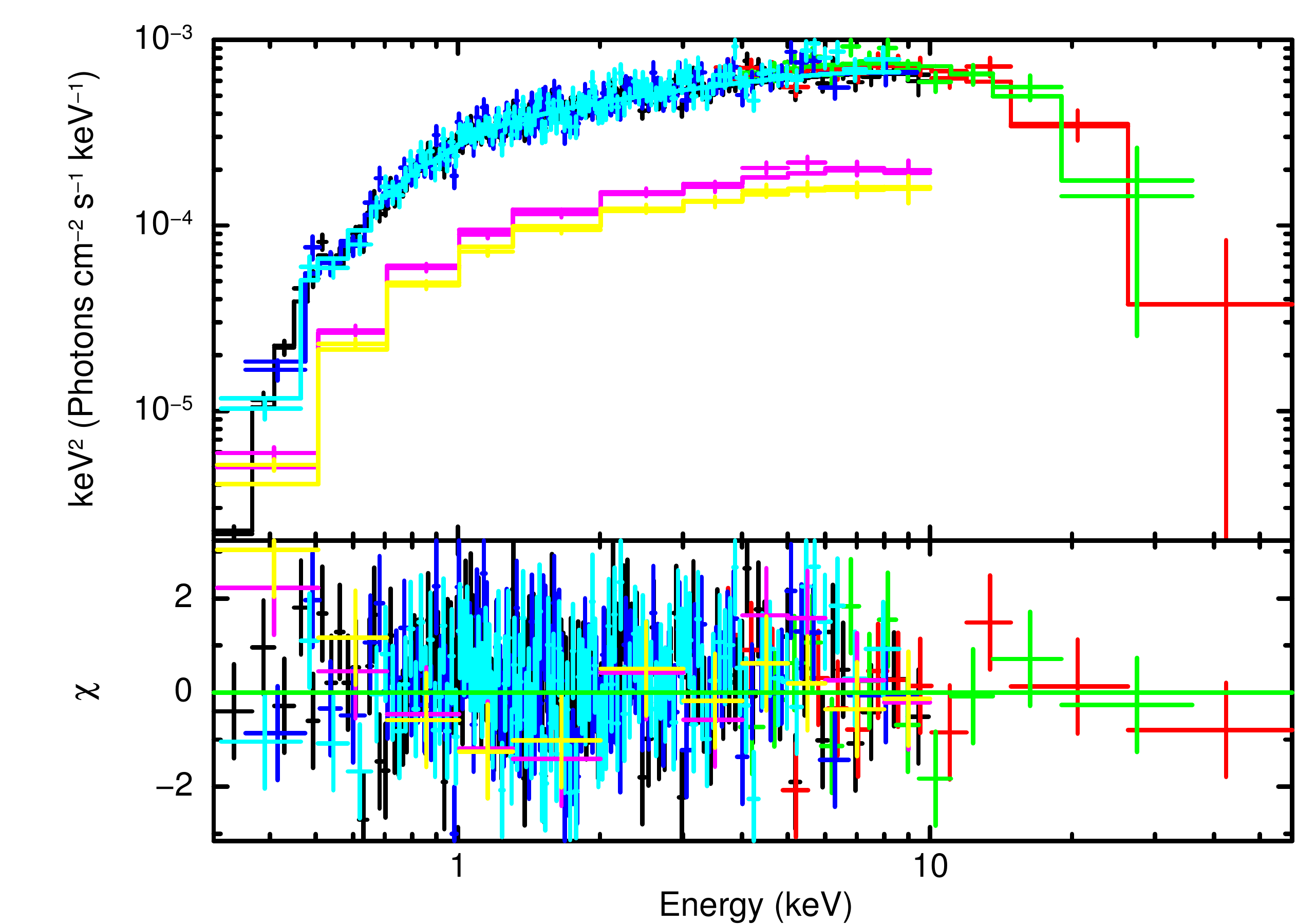}
   \caption{Top: unfolded $E^2f(E)$ spectra of ULX-1 fitted with an {\sc nthcomp} model (top), where the black, cyan, blue, red and green points are the EPIC-pn/MOS, NUSTAR FMPA and FMPB data, respectively. We add, in comparison, the CV spectra (see Sec.~\ref{tim}) where the fuchsia and yellow spectra are calculated for the timescales 300s--9600s and 1000s--16000s, respectively.}  
        \label{pnimage}
\end{figure}

\section{Data analysis and results}
\label{results}

\subsection{ULX-1}

\subsubsection{Spectral analysis}

The low counting statistics \chandra\ spectrum of ULX-1 can be modelled equally well with either an absorbed {\sc diskbb} or {\sc powerlaw} model. The {\sc powerlaw} gives a photon index $\Gamma=1.73\pm0.35$, a column density of N$_{\text{H}}=(0.4\pm0.2)\times10^{22}$ cm$^{-2}$ and a 0.3--10 keV absorbed flux of $(2.10\pm0.35)\times10^{-12}$ erg cm$^{-2}$ s$^{-1}$. 

Although this spectral shape is reminiscent of the {\it hard} state of XRBs \citep{mcclintock06}, such a possibility is ruled out when fitting the \xmm+\nus\ spectra. These show that the ULX-1 spectral properties are more complex than a simple {\sc diskbb} ($\chi^2/dof=2013.74/980$) or a {\sc powerlaw} ($\chi^2/dof=1060.91/980$, only marginally acceptable). We found that higher significance fits are indeed obtained with a modified accretion disc ({\sc diskpbb}), a powerlaw modified by an exponential cut-off ({\sc highecut$\times$powerlaw}), a Comptonization model ({\sc nthcomp}, \citealt{zdiarski96}) or a {\sc diskbb+bbody} (Table~\ref{table_spectra}). { However, the latter showed a degeneracy in the spectral parameters, allowing for both a low and high {\sc diskbb} temperature (see table~\ref{table_spectra}). The residuals of the cold-disc fit show a clear excess above 10 keV as observed in other ULXs \citep[see e.g.][]{walton18b}. On the other hand, the inner disc radius estimated from the fit with a {\sc diskpbb} model is $<6$ km (for an inclination angle $<60$\textdegree) which is unphysical.}

Focussing on the best fit with the {\sc nthcomp} model (Figure~\ref{pnimage}-top), which has the lowest $\chi^2_{\nu}$ value, we found a column density of $(0.24\pm0.05)\times10^{22}$ cm$^{-2}$, a photon index of $1.78\pm0.03$, and seed photons and electron temperatures of $0.15\pm0.04$ keV and $3.5\pm0.4$ keV, respectively. 
{ The seed photon temperature is quite low. Although not requested by the data, we added a {\sc diskbb} component fixing the temperature to that of the seed photons, as done in the past for a number of ULX spectra. The fit was, as expected, acceptable, with spectral parameters consistent with the single {\sc nthcomp} model, where the seed photon temperature converged to $0.2\pm0.05$ keV. Such a low temperature is consistent with the ones inferred from other bright ULXs when fitted with the adopted model \citep[e.g.][]{stobbart06,gladstone09,pintore12,pintore14,middleton15}, indicating that ULX-1 is not remarkably different from the rest of the bright ULX population. In addition, leaving the inner disc temperature free to vary, we also found a good fit with a seed photon and disc temperature of $0.5\pm0.2$ keV and $0.31\pm0.08$ keV, respectively. Because of the uncertainty in modeling the low energy part of the spectrum, the constraints on the seed photon temperature obtained with the {\sc nthcomp} model only should be treated with caution}.

The average absorbed 0.3--50 keV flux is $(2.50\pm0.06)\times10^{-12}$ erg cm$^{-2}$ s$^{-1}$, consistent with the average source flux during the \chandra\ observation. For a distance of 8.5 Mpc, the unabsorbed 0.3--50 keV luminosity is $\sim$2.7$\times$$10^{40}$ erg s$^{-1}$.

\subsubsection{Timing analysis}
\label{tim}

\begin{figure}
\center
		\includegraphics[width=8.9cm]{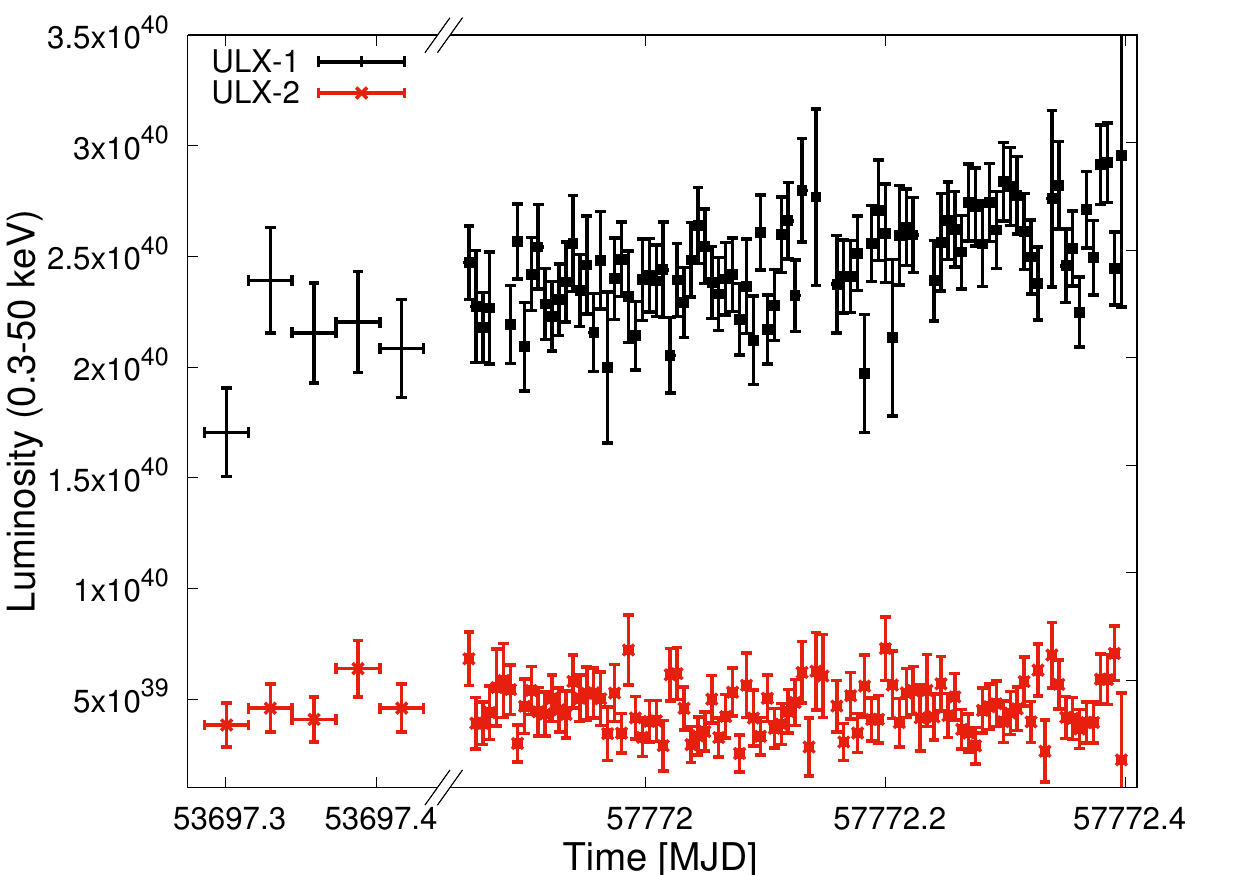}
		\includegraphics[width=8.9cm]{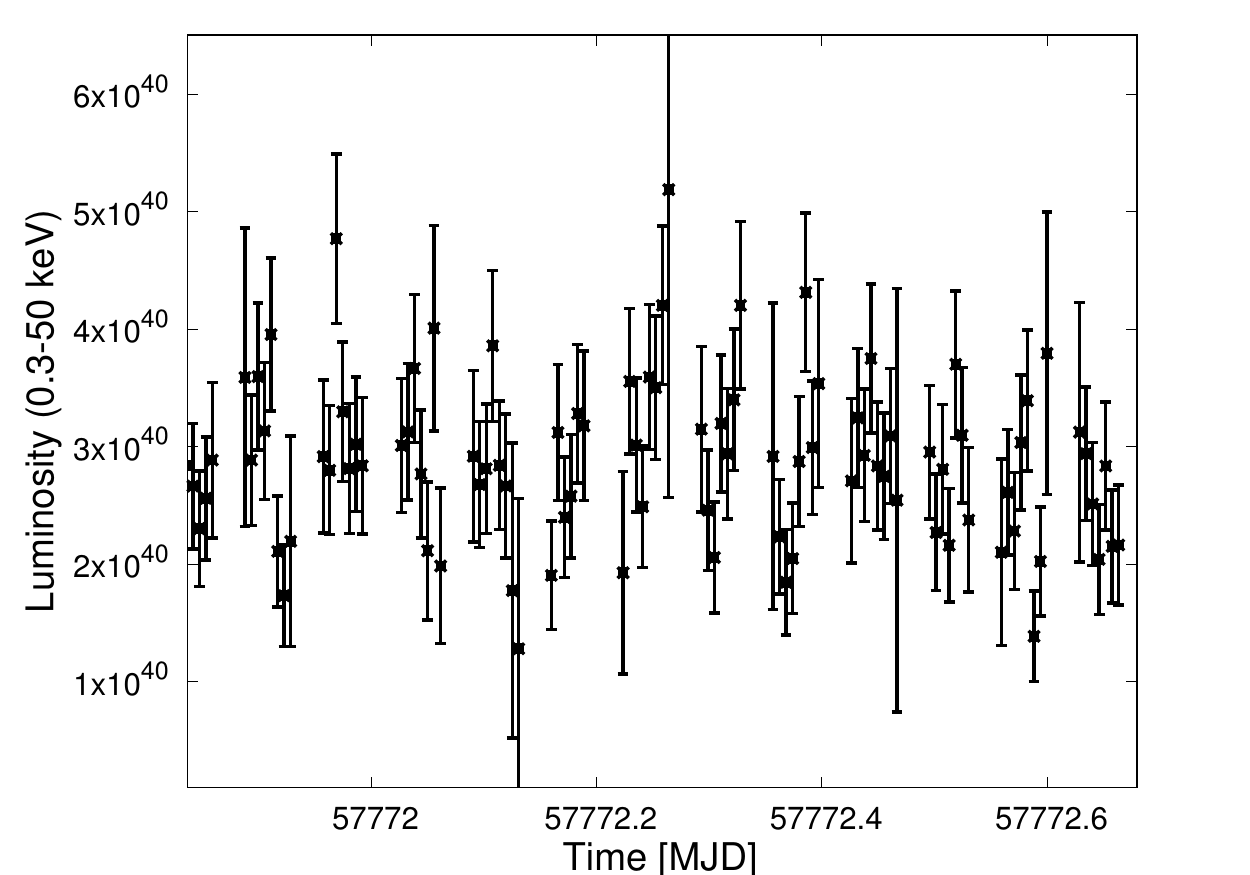}
		\vspace{-0.5cm}
   \caption{Top: \chandra\ (first 5 points) and { EPIC-pn} background-subtracted 0.3--50 keV lightcurve of ULX-1 (black) and ULX-2 (red), with a time binning of 500 s, { obtained assuming for both \xmm\ and \chandra\ datasets the best fits (Table~\ref{table_spectra}) with a {\sc nthcomp} and a {\sc bbody+diskbb} model, respectively (see main text)}. { Bottom: \nus\ (FMPA+FMPB) background-subtracted lightcurve, rebinned as the data in the top panel, and mostly overlapping in time with the EPIC-pn data.}}  
        \label{lc}
\end{figure}

The lightcurve of ULX-1 indicates an increment of $\sim30\%$ of the source luminosity during the \xmm\ and \nus\ observations. The \chandra\ lightcurve is instead rather constant and consistent with the flux at the beginning of the \xmm\ observation (see Figure~\ref{lc}). 
The EPIC lightcurve of ULX-1 shows also some high short-term variability. We splitted the 0.3--10 keV energy range in several intervals { (0.3--0.5, 0.5--0.7, 0.7--1.0, 1.0--1.3, 1.3--2.0, 2.0--3.0, 3.0--4.0, 4.0--5.0, 5.0--6.0, 6.0--7.0, 7.0--8.0 and 8.0--10 keV)} and for each of them we calculated the root mean square (RMS) variability. { We found that the RMS fractional variabilty (i.e. RMS divided by the average count rate of each band, e.g. \citealt{vaughan03}) is $\sim 45\%$ in each energy band. }

We further investigated such findings with the use of the covariance spectra \citep[CV; e.g.][]{wilkinson09,uttley14}, which have been proven as powerful tool to investigate the ULX timing properties \citep{middleton15}. We calculate the CV spectra in the same energy bands used for the RMS spectrum and adopting the 1.3--2.0 keV range as reference band. We choose two overlapping timescales to study short-term (300s--9600s) and long-term (1000s--16000s) correlated variability. As shown in Figure~\ref{pnimage}, both CV spectra follow the shape of the average spectrum. We fitted them with an {\sc nthcomp} model, where we fixed all the parameters, except the normalization, to the best-fit values of the average {\it XMM-Newton + NuSTAR} spectrum, and we found acceptable fits ($\chi^2/dof=14.62/11$ and $\chi^2/dof=15.93/11$, null hypothesis probability $>0.14$). Improvements in the best-fit statistics can be obtained by letting also the photon index and the seed photon temperature free to vary, but their uncertainties make the values still consistent with those of the average spectrum.
As we do not have any evidence or hint of a second spectral component in the CV spectra, these results imply that the variability is mainly driven by a single spectral component.

{ We also performed an accelerated search for coherent signals in the \xmm\ and \nus\ data, where we corrected
the arrival times of the events in a grid of about 600 $\dot{P} /P$ values in the range ($\pm$) $10^{-11} - 10^{-5}$ s s$^{-2}$. The search gave no statistically significant signals, and yielded a 3$\sigma$ upper limit on the fractional amplitude of 15\% in the period range 0.146s--100s.}

\begin{table*}
\center
          \caption{Spectral parameters for ULX-1 and ULX-2 in the \xmm+\nus. Errors at $90\%$  confidence level. The fits with $\chi^2/dof$ values in boldface are statistically acceptable (null hypothesis probability higher than 2$\sigma$).} 
      \label{table_spectra}
\scalebox{0.79}{\begin{minipage}{24cm}
\begin{tabular}{|l|c | ccc | cc | cccc | c|}
\hline
&  nH  & kT$_{disk}$ & $p$ & N$_{disk}$ & kT$_{bb}$/kT$_{seed}$ & N$_{bb}$ & $\Gamma$ & E$_{f}$ & E$_{c}$ & N$_{pow/nthcomp}$ & $\chi^2/dof$ \\
&  ($10^{22}$ cm$^{-2}$) & (keV) & & ($10^{-5}$) & (keV) & ($10^{-6}$) & & (keV) & (keV) &  ($10^{-5}$) &\\
\hline
ULX-1 & & & &&&&&&&&\\
\hline
{\sc diskbb} &  $0.079^{+0.006}_{-0.006}$ & $1.69^{+0.04}_{-0.04}$ & - & $1266^{+110}_{-98}$ & - & - & - & - & - & - & 2014/980\\
{\sc powerlaw} &  $0.32^{+0.01}_{-0.01}$ & - & - & - & - & - & $1.83^{+0.03}_{-0.03}$ & - & - & $45^{+1}_{-1}$ & 1061/980 \\
{\sc diskpbb} &  $0.30^{+0.01}_{-0.01}$ & $5.8^{+0.6}_{-0.5}$ & $0.531^{+0.004}_{-0.004}$ & $2.9^{+1.2}_{-1.0}$ & - & - & - & - & - & - & {\bf 970/979}\\
{\sc bbody+diskbb} &  $0.13^{+0.02}_{-0.02}$ & $2.9^{+0.2}_{-0.1}$ & - & $160^{+30}_{-30}$ & $0.34^{+0.02}_{-0.02}$ & $5.8^{+0.3}_{-0.3}$ & - & - & - & - & {\bf 1030/978}\\
{\sc bbody+diskbb alt.} &  $0.18^{+0.01}_{-0.01}$ & $0.74^{+0.04}_{-0.04}$ & - & $2.0^{+0.5}_{-0.4}\times10^4$ & $2.0^{+0.1}_{-0.1}$ & $18.7^{+0.8}_{-0.8}$ & - & - & - & - & 1068/978\\
{\sc highecut$\times$pow} &  $0.30^{+0.01}_{-0.01}$ & - & - & - & - & - & $1.78^{+0.03}_{-0.03}$ & $7.8^{+1.2}_{-1.4}$ & $13.4^{+4.1}_{-3}$ & $43^{+1}_{-1}$ & {\bf 962/978}\\
{\sc nthcomp} &  $0.24^{+0.04}_{-0.05}$ & - & - & - & $0.15^{+0.04}_{-0.04}$ & - & $1.78^{+0.03}_{-0.02}$ & - & $3.5^{+0.4}_{-0.3}$ & $39^{+3}_{-3}$ & {\bf 953/978}\\
\hline
\hline
ULX-2  & & & &&&&&&&&\\
\hline
{\sc diskbb} &  $0.06^{+0.02}_{-0.02}$ & $1.10^{+0.07}_{-0.07}$ & - & $830^{+230}_{-180}$ & - & - & - & - & - & - & 282/161\\
{\sc powerlaw} &  $0.30^{+0.03}_{-0.03}$ & - & - & - & - & - & $2.14^{+0.09}_{-0.09}$ & - & - & $8.0^{+0.7}_{-0.7}$ & {\bf 162/161} \\
{\sc diskpbb} &  $0.26^{+0.02}_{-0.02}$ & $3.1^{+1.0}_{-0.6}$ & $0.50^{+0.01}_{*}$ & $3.1^{+3.9}_{-3.1}$ & - & - & - & - & - & - & {\bf 166/160}\\
{\sc nthcomp} &  $0.21^{+0.1}_{-0.08}$ & - & - & - & $0.15^{+0.05}_{-0.15}$ & - & $2.07^{+0.10}_{-0.08}$ & - & 100 (fixed) & $7^{+1}_{-1}$ & {\bf 160/160}\\
{\sc bbody+diskbb} &  $0.23^{+0.05}_{-0.04}$ & $0.41^{+0.06}_{-0.05}$ & - & $3.5^{+3.0}_{-1.6}\times10^4$ & $1.2^{+0.1}_{-0.1}$& $2.0^{+0.2}_{-0.2}$ & - & - & - & - & {\bf 150/159}\\
\hline

\end{tabular}
\end{minipage}}
\end{table*}

\subsection{ULX-2}

\begin{figure}
\center
 		\includegraphics[angle=270,width=8.3cm]{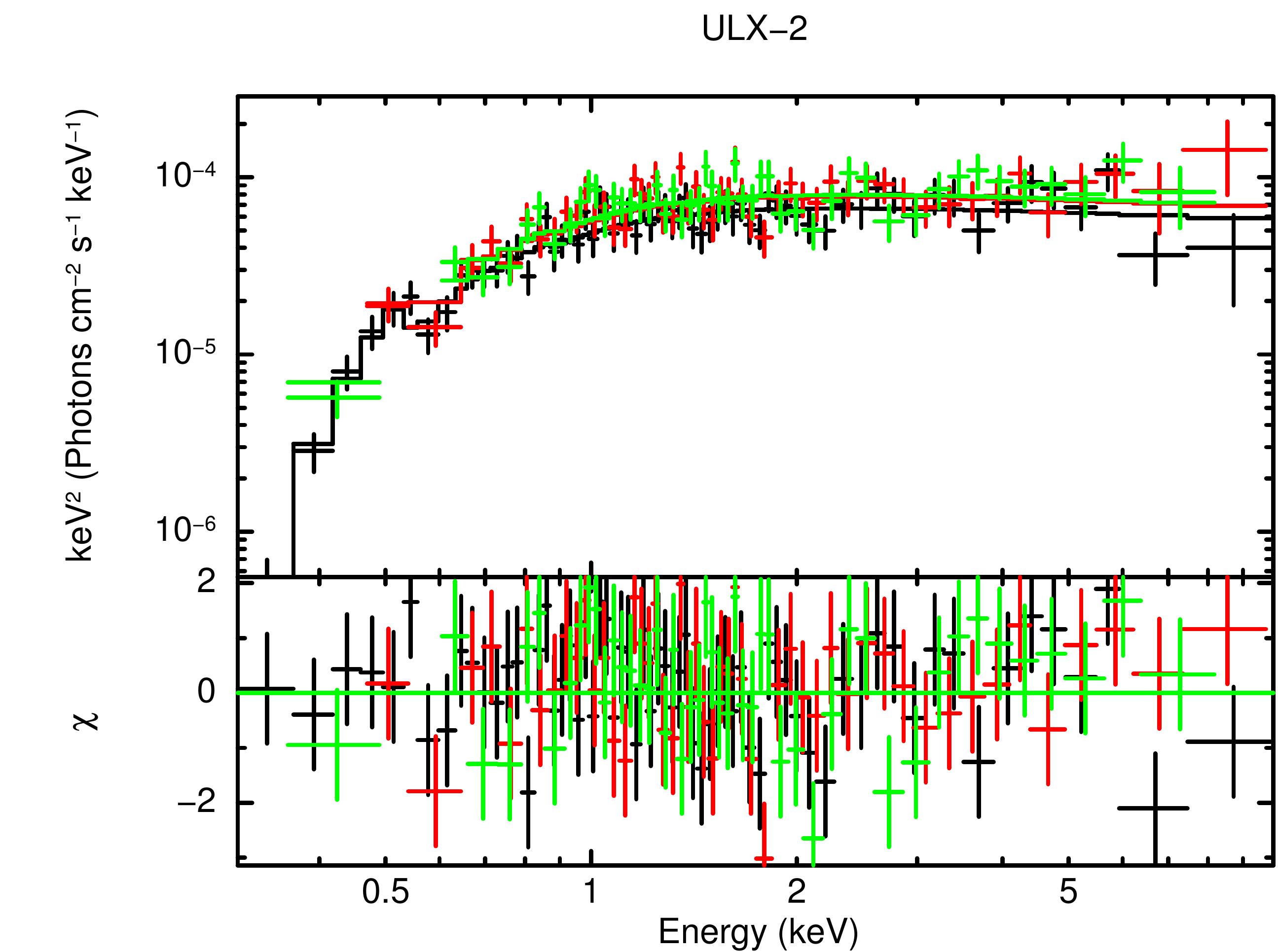}
  \caption{Unfolded $E^2f(E)$ spectra (top) and residuals (bottom) of ULX-2 when fitted with a {\sc powerlaw} model. The black, red and green spectra are the EPIC-pn and MOS1-2 data, respectively.}  
        \label{ulx2_spec}
\end{figure}

We found that ULX-2 showed a constant lightcurve during both \chandra\ and \xmm/\nus\ observations (Figure~\ref{lc}).

We firstly fitted its \chandra\ spectrum with an absorbed {\sc powerlaw}, where we  fixed the column density to $0.3\times10^{22}$ cm$^{-2}$ (see below). This model gives a photon index of $1.9\pm0.4$, again resembling a {\it hard} state of XRBs, and an absorbed 0.3--50 keV flux of $(7\pm1)\times10^{-13}$ erg cm$^{-2}$ s$^{-1}$. 

However, the \xmm\ data indicate that several models can provide statistically acceptable fits (see Table~\ref{table_spectra}), except for a single {\sc diskbb} model ($\chi^2/dof=282.3/161$). The {\sc powerlaw} model is characterized by a photon index of $\Gamma=2.14\pm0.09$ and a column density of $0.3\times10^{22}$ cm$^{-2}$ (Figure~\ref{ulx2_spec}). { We note that the ULX-2 \nus\ data are consistent with this spectral shape, although above 10 keV we found only upper limits.}
Instead, a fit with a {\sc diskbb+bbody} gives a column density of $0.23\times10^{22}$ cm$^{-2}$ and blackbody and multicolour blackbody disc temperatures of kT$_{bb}=1.2$ keV and kT$_{disc}=0.4$ keV, respectively. The corresponding 0.3--50 keV absorbed flux is $(4.0\pm0.15)\times10^{-13}$ erg cm$^{-2}$ s$^{-1}$. This implies an unabsorbed 0.3--50 keV luminosity of $\sim3.5\times10^{39}$ erg s$^{-1}$ (for a distance 8.5 Mpc). 

We did not find any significant short-term variability or coherent pulsation in the data. Adopting an accelerated search for coherent signals, we obtained upper limits on the fractional amplitude between $29\%-45\%$ in the period range 0.146s--100s.

\section{Discussion}
\label{discussion}

We have obtained the first high quality spectral data of the two ULXs in the galaxy NGC 925, which were considered promising IMBH candidates on the basis of archival low statistics \chandra\ spectra.

\subsection{ULX-2}
Unfortunately, we cannot draw firm conclusions on the nature of the compact object in ULX-2. In fact, its spectral properties are certainly not consistent with the thermal soft state of XRBs but, at the same time, it only marginally resembles a {\it hard} state (as the best-fit power-law may be too steep, $\Gamma\sim2.1$). 
We found that the ULX-2 spectrum could be also equally modelled by the combination of two thermal components (a multicolour disc plus a blackbody). Assuming an inclination angle $<60$\textdegree\ (because of the lack of dips or eclipses) and a distance of 8.5 Mpc, we estimate from the {\sc diskbb} normalization an inner disc radius of $\sim500-700$ km. Instead, a rough estimate of the {\sc bbody} emission gives an emitting radius of $73\pm13$ km. In the scenario of super-Eddington accretion, the former may be associated to the size of the region where outflows are ejected, while the latter might be the inner disc. Should the accreting compact object be a NS, these findings would allow us to exclude that the ULX-2 spectral components in the 0.3--10 keV energy band can be associated to the surface emission of the NS. 
Finally, we also note, from the optical observations, that ULX-2 is surrounded by a region of diffuse emission in H$_\alpha$ whose origin is not clear. Hence, further and deeper X-ray and optical observations of this source are strongly needed to better constrain its nature. 

\subsection{ULX-1}
\begin{figure}
\center
		\includegraphics[width=8.8cm]{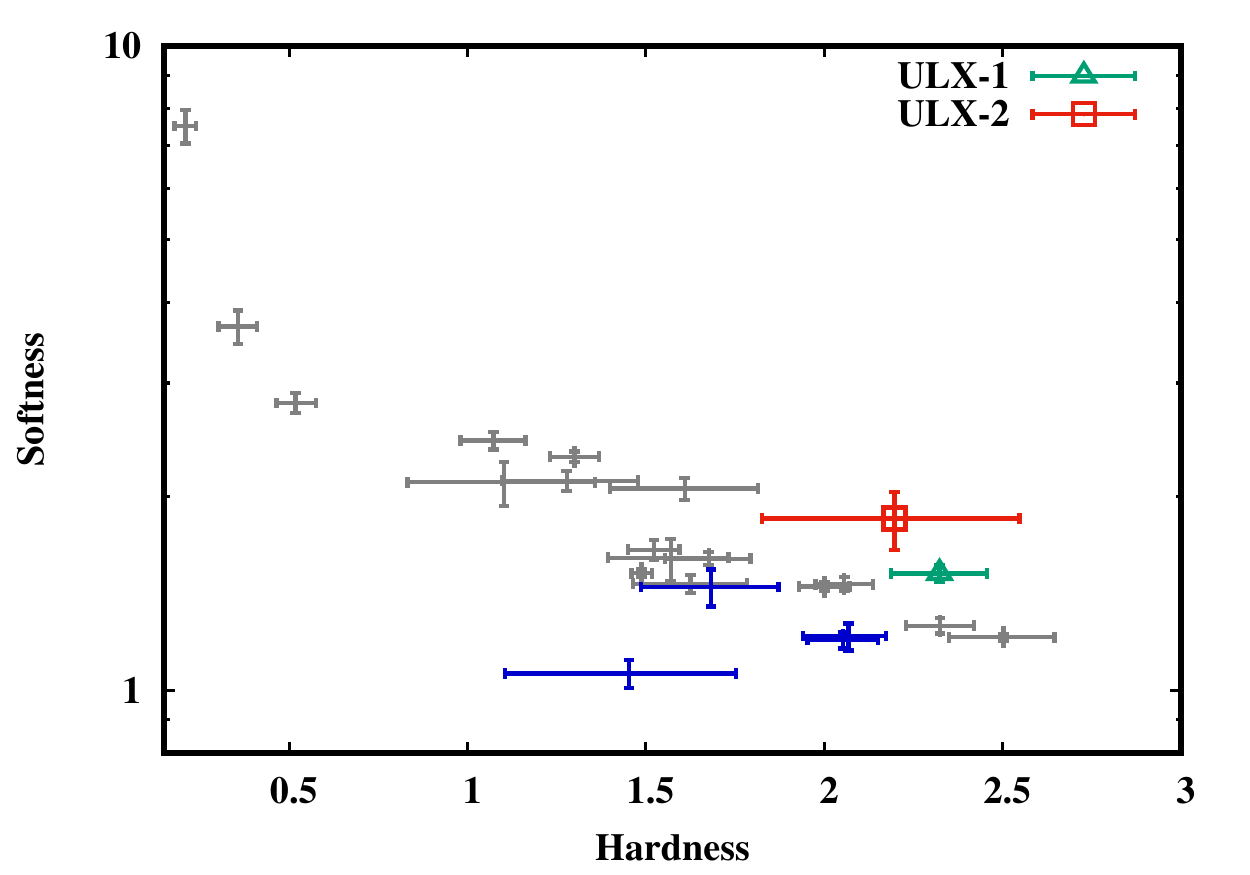}
		\vspace{-0.5cm}
   \caption{Color-color diagram obtained from the ratios of the fluxes in the energy ranges 2--4 keV, 4--6 keV and 6--30 keV and calculated from the best-fits with an absorbed {\sc highecut$\times$powerlaw}. The grey points are the sources shown in \citet{pintore17}, while the blue points are the two PULXs NGC 5907 X-1 and NGC 7793 P13. The green triangle and the red square indicate ULX-1 and ULX-2, respectively.}  
        \label{color-color}
\end{figure}

The high statistics obtained for ULX-1 allowed us some clearer insights on its nature. Assuming that the sub-Eddington accretion onto an IMBH shows the same states of the Galactic BH binaries, our results indicate that the IMBH scenario can be excluded and a super-Eddington accretion onto a stellar compact object appears more likely. 
In fact, its \xmm+\nus\ spectra revealed that the source has a very hard spectral shape but with a significant high energy cut-off. These spectral properties can be associated to a single optically thick Comptonization component, with electron temperature of $\sim$3.5 keV and seed photon temperature of 0.15 keV. Such a spectral shape allows us to classify ULX-1 as a {\it broadened disc} ULXs \citep{sutton13}. Its measured X-ray luminosity during the \xmm\ and \nus\ observations was $\sim2.5\times10^{40}$ erg s$^{-1}$. { However, analyzing the archival {\it Swift/XRT} observations\footnote{we used the online tool, http://www.swift.ac.uk/user\_objects/; \citet{evans09}} and assuming no significant spectral changes,} we found that ULX-1 reached a peak luminosity up to $\sim4\times10^{40}$ erg s$^{-1}$, making ULX-1 one the brightest known ULXs. We note that other sources with luminosities exceeding $10^{40}$ erg s$^{-1}$ present similar properties  \citep[e.g. NGC 470 ULX-1, Circinus ULX5, NGC 5907 X-1, NGC 5643 X-1][]{sutton12,walton13b,sutton13b,pintore16}.
Following \citet{pintore17}, we compared the position of ULX-1 with those of other well studied ULXs on a X-ray color-color diagram ({\it softness} = $(2-4$ keV$)/(4-6$ keV$)$, {\it hardness} = $(6-30$ keV$)/(4-6$ keV$)$). We found that the position of ULX-1 on this hardness-vs-softness diagram (Figure~\ref{color-color}) is close to that of the PULXs.
If we assume that hard ULX spectra may be associated to the PULXs, from this point of view alone we might consider ULX-1 as a possible candidate to host a NS.

{ Recently, \citet{walton18b} analyzed the spectra of a sample of bright ULXs adopting a model based on super-Eddington accretion onto magnetized NSs. They showed that all spectra leaved an excess above 10 keV well described by a cut-off powerlaw, the origin of which may be associated to the accretion column. This would suggest that all the ULXs of the sample can host NSs. In our analysis, we adopted a similar spectral model ({\sc diskbb+bbody}) and we also observed an excess at high energy, which can be modelled with a cut-off powerlaw (although not statistically requested by the data). Therefore, according to this spectral model, ULX-1 can be described as another NS accreting above Eddington, where its pulsed emission would not be detected because of the strong dilution by the high disc+wind emission. According to the work of \citet{koliopanos17}, who analyzed a similar ULX sample with a similar spectral model, it would be also possible to estimate the magnetic field of the NS in ULX-1: for a high blackbody temperature of $\sim2$ keV and considering a 0.3--10 keV unabsorbed luminosity of $\sim2\times10^{40}$ erg cm$^{-2}$ s$^{-1}$, the magnetic field should be $\sim5\times10^{13}$ G.}

ULX-1 showed also high short-term variability which rules out the scenario of a stable advection dominated disc (unless it was patchy; \citealt{miller14}). The RMS and covariance spectra showed that the strong variability is independent of the energy, supporting the hypothesis that it originates in a single component. Combining the ULX-1 spectral and temporal properties, we suggest that the source is seen at small inclination angles and, if the accretion rate is highly super-Eddington, strong (clumpy) outflows are ejected. MHD simulations \citep[e.g.][]{kawashima12} indicate that such outflows create a funnel around the central compact object. 
We probably see the hard X-ray emission produced in the regions very close to the compact object through such a funnel \citep[see e.g.][]{middleton15}.
In addition, the high level of short-term variability may be produced by the clumps of the outflow that occasionally cross our line of sight towards the central regions. The existence of powerful winds around ULX-1 is supported by the properties of its infrared counterpart: \citet{heida16} found that the IR emission is characterized by several emission features (where the Fe II is the dominant one), likely having their origin in an extended nebula around ULX-1 which is possibly fed by the ULX outflows. This interpretation is supported by our optical observations that show clear diffuse H$_{\alpha}$ emission around the source. Deeper observations in the optical band are definitely needed to constrain the origin of such a diffuse emission.

\section*{Acknowledgements} 

This work is partly based on observations collected at the Copernico telescope (Asiago, Italy) of the INAF-Osservatorio Astronomico di Padova, on observations obtained with \xmm, an ESA science mission with instruments and contributions directly funded by ESA Member States and NASA, and the NASA mission \nus.
FP and SM acknowledge the ``Contratto ASI -- INAF per analisi dati NuSTAR''.
PE acknowledges funding in the framework of the NWO Vidi award A.2320.0076. GR and GI Acknowledge that this research was supported in part through high performance computing resources and support provided by CINECA (MARCONI), awarded under the ISCRA initiative; and also through the INAF - CHIPP high performance computing project resources and support.

\addcontentsline{toc}{section}{Bibliography}
\bibliographystyle{mn2e}
\bibliography{biblio}

\begin{thebibliography}{}

\bibitem[\protect\citeauthoryear{{Arnaud}}{{Arnaud}}{1996}]{arnaud96}
{Arnaud} K.~A.,  1996, in {Jacoby}, G.~H. and {Barnes}, J., eds., Astronomical
  Data Analysis Software and Systems V. Vol.~101 of ASP Conf. Ser., San
  Francisco CA, {XSPEC: The First Ten Years}.
p.~17

\bibitem[\protect\citeauthoryear{{Bachetti}, {Harrison}, {Walton},
  {Grefenstette}, {Chakrabarty}, {F{\"u}rst}, {Barret} \& et al.}{{Bachetti}
  et~al.}{2014}]{bachetti14}
{Bachetti} M.,  {Harrison} F.~A.,  {Walton} D.~J.,  {Grefenstette} B.~W.,
  {Chakrabarty} D.,  {F{\"u}rst} F.,  {Barret}   et al. 2014, \nat, 514, 202

\bibitem[\protect\citeauthoryear{{Bachetti}, {Rana}, {Walton}, {Barret},
  {Harrison} \& et al.}{{Bachetti} et~al.}{2013}]{bachetti13}
{Bachetti} M.,  {Rana} V.,  {Walton} D.~J.,  {Barret} D.,  {Harrison} F.~A.,
  et al. 2013, \apj, 778, 163

\bibitem[\protect\citeauthoryear{{Caballero-Garcia}, {Belloni} \&
  {Wolter}}{{Caballero-Garcia} et~al.}{2013}]{caballero13}
{Caballero-Garcia} M.~D.,  {Belloni} T.,    {Wolter} A.,  2013, ArXiv e-prints

\bibitem[\protect\citeauthoryear{{Carpano}, {Haberl}, {Maitra} \&
  {Vasilopoulos}}{{Carpano} et~al.}{2018}]{carpano18a}
{Carpano} S.,  {Haberl} F.,  {Maitra} C.,    {Vasilopoulos} G.,  2018, \mnras

\bibitem[\protect\citeauthoryear{{Colbert} \& {Mushotzky}}{{Colbert} \&
  {Mushotzky}}{1999}]{colbert99}
{Colbert} E.~J.~M.,  {Mushotzky} R.~F.,  1999, \apj, 519, 89

\bibitem[\protect\citeauthoryear{{Evans}, {Beardmore}, {Page}, {Osborne},
  {O'Brien} \& et al.}{{Evans} et~al.}{2009}]{evans09}
{Evans} P.~A.,  {Beardmore} A.~P.,  {Page} K.~L.,  {Osborne} J.~P.,  {O'Brien}
  P.~T.,    et al. 2009, \mnras, 397, 1177

\bibitem[\protect\citeauthoryear{{Fabbiano}}{{Fabbiano}}{1989}]{fabbiano89}
{Fabbiano} G.,  1989, \araa, 27, 87

\bibitem[\protect\citeauthoryear{{Feng} \& {Soria}}{{Feng} \&
  {Soria}}{2011}]{fengsoria11}
{Feng} H.,  {Soria} R.,  2011, New Astronomy Reviews, 55, 166

\bibitem[\protect\citeauthoryear{{F{\"u}rst}, {Walton}, {Harrison}, {Stern},
  {Barret}, {Brightman}, {Fabian}, {Grefenstette}, {Madsen}, {Middleton},
  {Miller}, {Pottschmidt}, {Ptak}, {Rana} \& {Webb}}{{F{\"u}rst}
  et~al.}{2016}]{fuerst16a}
{F{\"u}rst} F.,  {Walton} D.~J.,  {Harrison} F.~A.,  {Stern} D.,  {Barret} D.,
  {Brightman} M.,  {Fabian} A.~C.,  {Grefenstette} B.,  {Madsen} K.~K.,
  {Middleton} M.~J.,  {Miller} J.~M.,  {Pottschmidt} K.,  {Ptak} A.,  {Rana}
  V.,    {Webb} N.,  2016, \apjl, 831, L14

\bibitem[\protect\citeauthoryear{{F{\"u}rst}, {Walton}, {Stern}, {Bachetti},
  {Barret}, {Brightman}, {Harrison} \& {Rana}}{{F{\"u}rst}
  et~al.}{2017}]{fuerst17}
{F{\"u}rst} F.,  {Walton} D.~J.,  {Stern} D.,  {Bachetti} M.,  {Barret} D.,
  {Brightman} M.,  {Harrison} F.~A.,    {Rana} V.,  2017, \apj, 834, 77

\bibitem[\protect\citeauthoryear{{Gladstone}, {Roberts} \& {Done}}{{Gladstone}
  et~al.}{2009}]{gladstone09}
{Gladstone} J.~C.,  {Roberts} T.~P.,    {Done} C.,  2009, \mnras, 397, 1836

\bibitem[\protect\citeauthoryear{{Heida}, {Jonker}, {Torres}, {Roberts},
  {Walton}, {Moon}, {Stern} \& {Harrison}}{{Heida} et~al.}{2016}]{heida16}
{Heida} M.,  {Jonker} P.~G.,  {Torres} M.~A.~P.,  {Roberts} T.~P.,  {Walton}
  D.~J.,  {Moon} D.-S.,  {Stern} D.,    {Harrison} F.~A.,  2016, \mnras, 459,
  771

\bibitem[\protect\citeauthoryear{{Heil}, {Vaughan} \& {Roberts}}{{Heil}
  et~al.}{2009}]{heil09}
{Heil} L.~M.,  {Vaughan} S.,    {Roberts} T.~P.,  2009, \mnras, 397, 1061

\bibitem[\protect\citeauthoryear{{Israel}, {Belfiore}, {Stella} \&
  {Esposito}}{{Israel} et~al.}{2017}]{israel16a}
{Israel} G.~L.,  {Belfiore} A.,  {Stella} L.,    {Esposito} P. e.~a.,  2017,
  Science, 355, 817

\bibitem[\protect\citeauthoryear{{Israel}, {Papitto}, {Esposito} \&
  {Stella}}{{Israel} et~al.}{2017}]{israel16b}
{Israel} G.~L.,  {Papitto} A.,  {Esposito} P.,    {Stella} L. e.~a.,  2017,
  \mnras, 466, L48

\bibitem[\protect\citeauthoryear{{Kaaret}, {Feng} \& {Roberts}}{{Kaaret}
  et~al.}{2017}]{kaaret17}
{Kaaret} P.,  {Feng} H.,    {Roberts} T.~P.,  2017, \araa, 55, 303

\bibitem[\protect\citeauthoryear{{Kawashima}, {Mineshige}, {Ohsuga} \&
  {Ogawa}}{{Kawashima} et~al.}{2016}]{kawashima16}
{Kawashima} T.,  {Mineshige} S.,  {Ohsuga} K.,    {Ogawa} T.,  2016, \pasj, 68,
  83

\bibitem[\protect\citeauthoryear{{Kawashima}, {Ohsuga}, {Mineshige}, {Yoshida},
  {Heinzeller} \& {Matsumoto}}{{Kawashima} et~al.}{2012}]{kawashima12}
{Kawashima} T.,  {Ohsuga} K.,  {Mineshige} S.,  {Yoshida} T.,  {Heinzeller} D.,
     {Matsumoto} R.,  2012, \apj, 752, 18

\bibitem[\protect\citeauthoryear{{Koliopanos}, {Vasilopoulos}, {Godet},
  {Bachetti}, {Webb} \& {Barret}}{{Koliopanos} et~al.}{2017}]{koliopanos17}
{Koliopanos} F.,  {Vasilopoulos} G.,  {Godet} O.,  {Bachetti} M.,  {Webb}
  N.~A.,    {Barret} D.,  2017, \aap, 608, A47

\bibitem[\protect\citeauthoryear{{Liu}, {Bregman}, {Bai}, {Justham} \&
  {Crowther}}{{Liu} et~al.}{2013}]{liu13}
{Liu} J.-F.,  {Bregman} J.~N.,  {Bai} Y.,  {Justham} S.,    {Crowther} P.,
  2013, \nat, 503, 500

\bibitem[\protect\citeauthoryear{{Madau} \& {Rees}}{{Madau} \&
  {Rees}}{2001}]{madau01}
{Madau} P.,  {Rees} M.~J.,  2001, \apjl, 551, L27

\bibitem[\protect\citeauthoryear{{Madsen}, {Beardmore}, {Forster}, {Guainazzi},
  {Marshall}, {Miller}, {Page} \& {Stuhlinger}}{{Madsen}
  et~al.}{2017}]{madsen17}
{Madsen} K.~K.,  {Beardmore} A.~P.,  {Forster} K.,  {Guainazzi} M.,  {Marshall}
  H.~L.,  {Miller} E.~D.,  {Page} K.~L.,    {Stuhlinger} M.,  2017, \aj, 153, 2

\bibitem[\protect\citeauthoryear{{McClintock} \& {Remillard}}{{McClintock} \&
  {Remillard}}{2006}]{mcclintock06}
{McClintock} J.~E.,  {Remillard} R.~A.,  2006, {in Compact stellar X-ray
  sources, ed. W. H. G. Levin and M. van der Klis}.
Cambridge: Cambridge University Press, p.~157

\bibitem[\protect\citeauthoryear{{Middleton}, {Heil}, {Pintore}, {Walton} \&
  {Roberts}}{{Middleton} et~al.}{2015}]{middleton15}
{Middleton} M.~J.,  {Heil} L.,  {Pintore} F.,  {Walton} D.~J.,    {Roberts}
  T.~P.,  2015, \mnras, 447, 3243

\bibitem[\protect\citeauthoryear{{Miller}, {Bachetti}, {Barret}, {Harrison},
  {Fabian}, {Webb}, {Walton} \& {Rana}}{{Miller} et~al.}{2014}]{miller14}
{Miller} J.~M.,  {Bachetti} M.,  {Barret} D.,  {Harrison} F.~A.,  {Fabian}
  A.~C.,  {Webb} N.~A.,  {Walton} D.~J.,    {Rana} V.,  2014, \apjl, 785, L7

\bibitem[\protect\citeauthoryear{{Miller} \& {Hamilton}}{{Miller} \&
  {Hamilton}}{2002}]{miller02}
{Miller} M.~C.,  {Hamilton} D.~P.,  2002, \mnras, 330, 232

\bibitem[\protect\citeauthoryear{{Mizuno}, {Miyawaki}, {Ebisawa} \& et
  al.}{{Mizuno} et~al.}{2007}]{mizuno07}
{Mizuno} T.,  {Miyawaki} R.,  {Ebisawa} K.,    et al. 2007, \pasj, 59, 257

\bibitem[\protect\citeauthoryear{{Motch}, {Pakull}, {Soria}, {Gris{\'e}} \&
  {Pietrzy{\'n}ski}}{{Motch} et~al.}{2014}]{motch14}
{Motch} C.,  {Pakull} M.~W.,  {Soria} R.,  {Gris{\'e}} F.,    {Pietrzy{\'n}ski}
  G.,  2014, \nat, 514, 198

\bibitem[\protect\citeauthoryear{{Mushtukov}, {Suleimanov}, {Tsygankov} \&
  {Ingram}}{{Mushtukov} et~al.}{2017}]{mushtukov17}
{Mushtukov} A.~A.,  {Suleimanov} V.~F.,  {Tsygankov} S.~S.,    {Ingram} A.,
  2017, \mnras, 467, 1202

\bibitem[\protect\citeauthoryear{{Mushtukov}, {Suleimanov}, {Tsygankov} \&
  {Poutanen}}{{Mushtukov} et~al.}{2015}]{mushtukov15}
{Mushtukov} A.~A.,  {Suleimanov} V.~F.,  {Tsygankov} S.~S.,    {Poutanen} J.,
  2015, \mnras, 454, 2539

\bibitem[\protect\citeauthoryear{{Ohsuga}, {Mineshige}, {Mori} \&
  {Kato}}{{Ohsuga} et~al.}{2009}]{ohsuga09}
{Ohsuga} K.,  {Mineshige} S.,  {Mori} M.,    {Kato} Y.,  2009, \pasj, 61, L7

\bibitem[\protect\citeauthoryear{{Pinto}, {Middleton} \& {Fabian}}{{Pinto}
  et~al.}{2016}]{pinto16}
{Pinto} C.,  {Middleton} M.~J.,    {Fabian} A.~C.,  2016, \nat, 533, 64

\bibitem[\protect\citeauthoryear{{Pintore} \& {Zampieri}}{{Pintore} \&
  {Zampieri}}{2012}]{pintore12}
{Pintore} F.,  {Zampieri} L.,  2012, \mnras, 420, 1107

\bibitem[\protect\citeauthoryear{{Pintore}, {Zampieri}, {Stella}, {Wolter},
  {Mereghetti} \& {Israel}}{{Pintore} et~al.}{2017}]{pintore17}
{Pintore} F.,  {Zampieri} L.,  {Stella} L.,  {Wolter} A.,  {Mereghetti} S.,
  {Israel} G.~L.,  2017, \apj, 836, 113

\bibitem[\protect\citeauthoryear{{Pintore}, {Zampieri}, {Sutton}, {Roberts},
  {Middleton} \& {Gladstone}}{{Pintore} et~al.}{2016}]{pintore16}
{Pintore} F.,  {Zampieri} L.,  {Sutton} A.~D.,  {Roberts} T.~P.,  {Middleton}
  M.~J.,    {Gladstone} J.~C.,  2016, \mnras, 459, 455

\bibitem[\protect\citeauthoryear{{Pintore}, {Zampieri}, {Wolter} \&
  {Belloni}}{{Pintore} et~al.}{2014}]{pintore14}
{Pintore} F.,  {Zampieri} L.,  {Wolter} A.,    {Belloni} T.,  2014, \mnras,
  439, 3461

\bibitem[\protect\citeauthoryear{{Portegies Zwart}, {Baumgardt}, {Hut},
  {Makino} \& {McMillan}}{{Portegies Zwart} et~al.}{2004}]{portegies04}
{Portegies Zwart} S.~F.,  {Baumgardt} H.,  {Hut} P.,  {Makino} J.,
  {McMillan} S.~L.~W.,  2004, \nat, 428, 724

\bibitem[\protect\citeauthoryear{{Poutanen}, {Lipunova}, {Fabrika}, {Butkevich}
  \& {Abolmasov}}{{Poutanen} et~al.}{2007}]{poutanen07}
{Poutanen} J.,  {Lipunova} G.,  {Fabrika} S.,  {Butkevich} A.~G.,
  {Abolmasov} P.,  2007, \mnras, 377, 1187

\bibitem[\protect\citeauthoryear{{Rana}, {Harrison}, {Bachetti}, {Walton},
  {Furst}, {Barret}, {Miller}, {Fabian}, {Boggs}, {Christensen}, {Craig},
  {Grefenstette}, {Hailey}, {Madsen}, {Ptak}, {Stern}, {Webb} \&
  {Zhang}}{{Rana} et~al.}{2014}]{rana14}
{Rana} V.,  {Harrison} F.~A.,  {Bachetti} M.,  {Walton} D.~J.,  {Furst} F.,
  {Barret} D.,  {Miller} J.~M.,  {Fabian} A.~C.,  {Boggs} S.~E.,  {Christensen}
  F.~C.,  {Craig} W.~W.,  {Grefenstette} B.~W.,  {Hailey} C.~J.,  {Madsen}
  K.~K.,  {Ptak} A.~F.,  {Stern} D.,  {Webb} N.~A.,    {Zhang} W.~W.,  2014,
  ArXiv e-prints

\bibitem[\protect\citeauthoryear{{Roberts}}{{Roberts}}{2007}]{roberts07}
{Roberts} T.~P.,  2007, \apss, 311, 203

\bibitem[\protect\citeauthoryear{{Stobbart}, {Roberts} \& {Wilms}}{{Stobbart}
  et~al.}{2006}]{stobbart06}
{Stobbart} A.-M.,  {Roberts} T.~P.,    {Wilms} J.,  2006, \mnras, 368, 397

\bibitem[\protect\citeauthoryear{{Sutton}, {Roberts}, {Gladstone}, {Farrell},
  {Reilly}, {Goad} \& {Gehrels}}{{Sutton} et~al.}{2013}]{sutton13b}
{Sutton} A.~D.,  {Roberts} T.~P.,  {Gladstone} J.~C.,  {Farrell} S.~A.,
  {Reilly} E.,  {Goad} M.~R.,    {Gehrels} N.,  2013, \mnras, 434, 1702

\bibitem[\protect\citeauthoryear{{Sutton}, {Roberts} \& {Middleton}}{{Sutton}
  et~al.}{2013}]{sutton13}
{Sutton} A.~D.,  {Roberts} T.~P.,    {Middleton} M.~J.,  2013, \mnras, 435,
  1758

\bibitem[\protect\citeauthoryear{{Sutton}, {Roberts}, {Walton}, {Gladstone} \&
  {Scott}}{{Sutton} et~al.}{2012}]{sutton12}
{Sutton} A.~D.,  {Roberts} T.~P.,  {Walton} D.~J.,  {Gladstone} J.~C.,
  {Scott} A.~E.,  2012, \mnras, 423, 1154

\bibitem[\protect\citeauthoryear{{Swartz}, {Soria}, {Tennant} \&
  {Yukita}}{{Swartz} et~al.}{2011}]{swartz11}
{Swartz} D.~A.,  {Soria} R.,  {Tennant} A.~F.,    {Yukita} M.,  2011, \apj,
  741, 49

\bibitem[\protect\citeauthoryear{{Takeuchi}, {Ohsuga} \&
  {Mineshige}}{{Takeuchi} et~al.}{2013}]{takeuchi13}
{Takeuchi} S.,  {Ohsuga} K.,    {Mineshige} S.,  2013, \pasj, 65, 88

\bibitem[\protect\citeauthoryear{{Uttley}, {Cackett}, {Fabian}, {Kara} \&
  {Wilkins}}{{Uttley} et~al.}{2014}]{uttley14}
{Uttley} P.,  {Cackett} E.~M.,  {Fabian} A.~C.,  {Kara} E.,    {Wilkins} D.~R.,
   2014, \aapr, 22, 72

\bibitem[\protect\citeauthoryear{{Vaughan}, {Edelson}, {Warwick} \&
  {Uttley}}{{Vaughan} et~al.}{2003}]{vaughan03}
{Vaughan} S.,  {Edelson} R.,  {Warwick} R.~S.,    {Uttley} P.,  2003, \mnras,
  345, 1271

\bibitem[\protect\citeauthoryear{{Walton}, {Fuerst}, {Harrison}, {Stern},
  {Bachetti}, {Barret}, {Bauer}, {Boggs}, {Christensen}, {Craig}, {Fabian},
  {Grefenstette}, {Hailey}, {Madsen}, {Miller}, {Ptak}, {Rana}, {Webb} \&
  {Zhang}}{{Walton} et~al.}{2013}]{walton13b}
{Walton} D.~J.,  {Fuerst} F.,  {Harrison} F.,  {Stern} D.,  {Bachetti} M.,
  {Barret} D.,  {Bauer} F.,  {Boggs} S.~E.,  {Christensen} F.~E.,  {Craig}
  W.~W.,  {Fabian} A.~C.,  {Grefenstette} B.~W.,  {Hailey} C.~J.,  {Madsen}
  K.~K.,  {Miller} J.~M.,  {Ptak} A.,  {Rana} V.,  {Webb} N.~A.,    {Zhang}
  W.~W.,  2013, \apj, 779, 148

\bibitem[\protect\citeauthoryear{{Walton}, {F{\"u}rst}, {Harrison}, {Stern},
  {Bachetti}, {Barret}, {Brightman}, {Fabian}, {Middleton}, {Ptak} \&
  {Tao}}{{Walton} et~al.}{2018}]{walton18}
{Walton} D.~J.,  {F{\"u}rst} F.,  {Harrison} F.~A.,  {Stern} D.,  {Bachetti}
  M.,  {Barret} D.,  {Brightman} M.,  {Fabian} A.~C.,  {Middleton} M.~J.,
  {Ptak} A.,    {Tao} L.,  2018, \mnras, 473, 4360

\bibitem[\protect\citeauthoryear{{Walton}, {F{\"u}rst}, {Heida}, {Harrison},
  {Barret}, {Stern}, {Bachetti}, {Brightman}, {Fabian} \& {Middleton}}{{Walton}
  et~al.}{2018}]{walton18b}
{Walton} D.~J.,  {F{\"u}rst} F.,  {Heida} M.,  {Harrison} F.~A.,  {Barret} D.,
  {Stern} D.,  {Bachetti} M.,  {Brightman} M.,  {Fabian} A.~C.,    {Middleton}
  M.~J.,  2018, \apj, 856, 128

\bibitem[\protect\citeauthoryear{{Walton}, {Harrison}, {Grefenstette}, {Miller}
  \& et al.}{{Walton} et~al.}{2014}]{walton14}
{Walton} D.~J.,  {Harrison} F.~A.,  {Grefenstette} B.~W.,  {Miller} J.~M.,
  et al. 2014, \apj, 793, 21

\bibitem[\protect\citeauthoryear{{Walton}, {Miller}, {Harrison}, {Fabian},
  {Roberts}, {Middleton} \& {Reis}}{{Walton} et~al.}{2013}]{walton13}
{Walton} D.~J.,  {Miller} J.~M.,  {Harrison} F.~A.,  {Fabian} A.~C.,  {Roberts}
  T.~P.,  {Middleton} M.~J.,    {Reis} R.~C.,  2013, \apjl, 773, L9

\bibitem[\protect\citeauthoryear{{Wilkinson} \& {Uttley}}{{Wilkinson} \&
  {Uttley}}{2009}]{wilkinson09}
{Wilkinson} T.,  {Uttley} P.,  2009, \mnras, 397, 666

\bibitem[\protect\citeauthoryear{{Zampieri} \& {Roberts}}{{Zampieri} \&
  {Roberts}}{2009}]{zampieri09}
{Zampieri} L.,  {Roberts} T.~P.,  2009, \mnras, 400, 677

\bibitem[\protect\citeauthoryear{{Zdziarski}, {Johnson} \&
  {Magdziarz}}{{Zdziarski} et~al.}{1996}]{zdiarski96}
{Zdziarski} A.~A.,  {Johnson} W.~N.,    {Magdziarz} P.,  1996, \mnras, 283, 193

\end{thebibliography}

\bsp
\label{lastpage}
\end{document}